\documentclass[12pt]{iopart}

\begin{document}

\title[Entangling capabilities of symmetric two qubit gates]{Entangling capabilities of symmetric two qubit gates}

\author{Swarnamala\,\,  Sirsi, Veena Adiga\footnote{Also at : St.Joseph's College(autonomous), Bengaluru -27,\,India} and Subramanya Hegde}

\address{Yuvaraja's College, 
University of Mysore, Mysore-05,\,India}
\ead{vadiga11@gmail.com}

\begin{abstract}
Our work addresses the problem of generating maximally entangled two spin-1/2 (qubit) symmetric states using NMR, NQR, Lipkin-Meshkov-Glick 
Hamiltonians. Time evolution of such Hamiltonians provides various logic gates which can be used for quantum processing tasks. 
Pairs of spin-1/2's have modeled a wide range of problems in physics. Here we are interested in two spin-1/2 symmetric states which belong to a 
subspace spanned by the angular  momentum basis $\{| j=1, \mu\rangle; \mu= +1,0,-1\}$. Our technique relies on the decomposition of a 
Hamiltonian in terms of SU(3) generators. In this context, we define a set of linearly independent, traceless, Hermitian operators which provides an 
alternate set of SU(n) generators. These matrices are constructed out of angular momentum operators $\bf{J}_x$,$\bf{J}_y$,$\bf{J}_z$.  We construct 
and study the properties of perfect entanglers acting on a symmetric subspace i.e., spin-1 operators that can generate maximally entangled states from some 
suitably chosen initial separable states in terms of their entangling power. 

\end{abstract}

%\noindent{\it Keywords}: 
\maketitle

\section{Introduction}
In the last few years there has been considerable increase in experimental activity \cite{Fortschr} aiming to create entangled quantum states 
which have potential applications in quantum information processing tasks. In practice, these states are created by some physical operations involving
the interaction between several systems. Thus analyzing these operations with regard to the possibility of creating maximally entangled states from an initial 
unentangled one and characterization of entangling capabilities of quantum operators play an important role in quantum information theory.  
Spin-1/2 (two level) systems have modeled a wide range of problems in physics. Considering N spin-1/2's (N qubits) in the symmetric subspace -- 
the set of those N-particle pure states that remain unchanged by permutations of individual particles \cite{Werner,Wang},   
we define a set of $(N+1)^{2} $ linearly independent, experimentally realizable cartesian tensor operators which provide different logic gates for quantum 
computation. In the particular case of two spin-1/2's NMR , NQR provide "hardware" for realizable quantum computers which involve the study of time
evolution of Hamiltonian which can also be  time dependent. Since these two qubit symmetric gates are capable of producing entanglement, quantifying their entangling capability is very important. Makhlin \cite {Makhlin} has analyzed nonlocal properties of two-qubit gates
and also studied some basic properties of perfect entanglers which are defined as the unitary operators that can generate maximally entangled states 
from some suitably chosen separable states. Zanardi et al. \cite {Zanardi} have explored the entangling power of quantum evolutions in terms of mean 
linear entropy produced when unitary operator acts on a given distribution of pure product states. Kraus and Cirac \cite{Kraus}, Rezakhani \cite {Rezakhani}
have given the tools to find the best separable two qubit input orthonormal product states such that some given unitary transformation can create 
maximally entangled quantum states. The entangling capability of a unitary quantum gate can be quantified by its  entangling power $ e_{p}(U)$
\cite{Zanardi}. Balakrishnan et al. \cite{Balakrishnan} have derived $ e_{p}(U)$ in terms of local invariant $G_{1}$. 
In this paper, we show that the two qubit symmetric quantum gates expressed in terms of newly defined linearly independent cartesian tensor operators
belong to the class of perfect entanglers which can generate maximally entangled states from some suitably chosen product states. Further we show that
these symmetric two qubit gates belong to a family of special perfect entanglers under certain conditions. This is a very relevant problem not only 
from the theoretical point of view, but also from the experimental one.

In section 2, we define an alternate representation of SU(n) generators using the well known spherical tensor operators. Explicit form of these 
generators in 3-dimensional representation is given in 2.1. Section 2.2 deals with algebra of these cartesian tensor operators. Section 3 deals
with two qubit symmetric gates and their entangling power in terms of the invariant $G_{1}$. In section 3.1, we identify the conditions under which
the perfect entangler can be classified as special perfect entangler. Further the entangling property of Lipkin-Meshkov-Glick Hamiltonian is studied
in the spin-1 subspace.   

\subsection{\bf{Symmetric states}}
Symmetric states offer elegant mathematical analysis as the dimension of the Hilbert space reduces drastically from $2^N$ to (N + 1), when N qubits respect exchange symmetry. Such a 
Hilbert space is considered to be spanned by the eigen states $\{| j,m\rangle; -j\geq m\leq+j\}$  of angular  momentum operators $J^2$ and $J_z$ , 
where $j= \frac{N}{2}$. The corresponding density matrix gets transformed to a $3\times 3$ block form in the symmetric subspace charactrized by the
maximal value of total angular momentum $j_{max} = 1$. The symmetric subspace  provides a convenient, albeit idealized, computationally accessible 
class of spin states relevant to many experimental situations such as spin squeezing. Completely symmetric systems are experimentally interesting, 
largely because it is often easier to nonselectively address an entire ensemble of particles rather than individually address each member.
Permutationally symmetric states are useful in a variety of quantum information processing tasks and a class of these states have recently been 
implemented experimentally \cite{Zeilinger,Toth}.

\section{\bf{Alternative representation of SU(n) generators}} 
It is well known that any Hermitian operator for a spin j system is given by \cite{Fano} 
\begin{equation}
\mathcal {H}(\vec J) = \frac {1}{(2j+1)}\sum^{2j}_{k=0}\,\sum^{+k}_{q=-k}\,\, h^{k}_{q}\, \tau^{k^{\dagger}}_{q}({\vec J}),
\end{equation}
where $\tau^{k}_{q}\,{'}s$ \, (with $\tau^{0}_{0} = I$ , the identity operator) are irreducible spherical tensor operators of rank `k' in the 2j+1 
dimension spin space with projection `q' along the axis of quantization in the real 3-dimensional space. The $\tau^{k}_{q}\,{'}s$ satisfy the 
orthogonality relation
\begin{equation}
Tr({\tau^{k^{\dagger}}_{q}\tau^{k^{'}}_{q^{'}}})= (2j+1)\,\delta_{kk^{'}} \delta_{qq^{'}}\,.
\end{equation}
Here the normalization has been chosen so as to be in agreement with Madison convention \cite {Satchler}. The spherical tensor parameters $h^{k}_{q}$ which 
characterize the given Hermitian operator $\mathcal {H}$ are given by $h^{k}_{q} = {Tr({\mathcal {H}\,\tau^{k}_{q}})}$.
Since $\mathcal {H}$ is Hermitian and $\tau^{k^{\dagger}}_{q} = (-1)^{q}\tau^{k}_{-q}$\,\,,  $h^{k}_{q}\,{'}s$ satisfy the condition 
$h^{k^{*}}_{q} = (-1)^{q}\,h^{k}_{-q}$.
The spherical tensor parameters ${h^{k}_{q}}\,{'}s$ have simple transformation properties under co-ordinate rotation \cite{Rose} in the 3-dimensional space. i.e.,
\begin{equation}
(h^{k}_{q})^{R} = \sum^{+k}_{q^{'}=-k}\,\, D^{k}_{q^{'}q}(\alpha\beta\gamma)\,h^{k}_{q^{'}}\,\,,
\end{equation}
where $D^{k}_{q^{'}q}(\alpha\beta\gamma)$ denote Wigner-D matrix parametrized by Euler angles $(\alpha\beta\gamma)$.

Following  the well known Weyl construction \cite{Rose} for $\tau^{k}_{q}$  in terms of angular momentum operators $J_{x}$, $J_{y}$ and $J_{z}$, we have   
\begin{equation}
\tau_{q}^{k}(\vec{J}) =  \mathcal {N}_{kj}\,(\vec{J}\cdot \vec{\bigtriangledown})^k \,r^{k} \,{Y}^{k}_{q}(\hat{r})\,,
\end{equation} 
where 
\begin{equation}\mathcal {N}_{kj}=\frac{2^{k}}{k!}\sqrt{\frac{4\pi(2j-k)!(2j+1)}{(2j+k+1)!}}, 
\end{equation} are the normalization factors and ${Y}^{k}_{q}(\hat {r})$ are the spherical harmonics.
Under rotations $\tau^{k}_{q}\,{'}s$ transform according to Wigner-D matrices. i.e.,
\begin{equation}
(\tau^{k}_{q}(\vec{J}))^{R} = \sum^{2j}_{k=0}\sum^{+k}_{q=-k}\,\, D^{k}_{q^{'}q}(\alpha\beta\gamma)\,\tau^{k}_{q^{'}}(\vec{J})\,\,.
\end{equation}

We now define a set of linearly independent, traceless (except $(T^{0})^{0}_{0}$), orthonormal Hermitian basis matrices 
$ (T^{\alpha})_{q}^{k}$ \,\,, where $\alpha = +,-,0$ , $k = 1...2j$\,\,, and  q = 1 to +k as follows:
\begin{equation}
(T^{+})_{q}^{k} = \frac {\tau^{k}_{q} + (\tau^{k}_{q})^\dagger}{\sqrt {2(2j+1)}}\,\,,
\end{equation} 
\begin{equation}
(T^{-})_{q}^{k} = \frac {i (\tau^{k}_{q} - (\tau^{k}_{q})^\dagger)}{\sqrt {2(2j+1)}}\,\,\,, 
\end{equation} 
and 
\begin{equation}
(T^{0})_{0}^{k} = \frac {\tau^{k}_{0}}{\sqrt {2j+1}}\,\,\,.
\end{equation} 
Observe that these matrices satisfy the relation $Tr((T^{\alpha})_{q}^{k}(T^{\beta})_{q'}^{k'})$ = $\delta_{\alpha\beta}\delta_{kk^{'}}\delta_{qq^{'}}$.
In our new  representation the most general density matrix can be written as
\begin{equation}
\rho =  (r^{0})^{0}_{0}(T^{0})^{0}_{0}+\sum_{k = 1....2j}(r^{0})^{k}_{0}(T^{0})^{k}_{0}+\sum_{\alpha = +,- }\sum_{k=1...2j}\sum_{q=1...k}(r^{\alpha})^{k}_{q}(T^{\alpha})^{k}_{q}
\end{equation}
Apart from $(T^{0})^{0}_{0}$ which is proportional to identity matrix, there are 2j diagonal matrices namely $(T^{0})^{k}_{0}$\,,\,k = 1...2j and 
the rest are off diagonal.
 
\subsection{SU(3) generators:}
In the particular case of two qubit symmetric subspace, our set of basis matrices \footnote{The above matrices are equivalent to the set 
of matrices with different normalization defined by R.J.Morris \cite{Morris}.} can be obtained from Equation(7,8,9) as \footnote{We have used a 
different notation for the basis set for the sake of simplicity.}

\begin{eqnarray}
 \eqalign M_{0}= \sqrt{\frac{2}{3}}\tau^{0}_{0}~~, \qquad\mbox M_{1}= {\frac{\tau^{1}_{1}+\tau^{1\dagger}_{1}}{\sqrt{3}}}~~,\qquad\mbox M_{2}= \frac {i(\tau^{1}_{1}-\tau^{1\dagger}_{1})}{\sqrt{3}}\,,\cr
  \eqalign M_{3}= \sqrt{\frac{2}{3}}\tau^{1}_{0}~,\qquad\mbox M_{4}= \frac {i(\tau^{2}_{2}-\tau^{2\dagger}_{2})}{\sqrt{3}}~,\qquad\mbox M_{5}=\frac {i(\tau^{2}_{1}-\tau^{2\dagger}_{1})}{\sqrt{3}}~,\label{eq2} \cr
\eqalign M_{6}=\frac{\tau^{2}_{1}+\tau^{2\dagger}_{1}}{\sqrt{3}}~,\qquad\mbox M_{7}=\frac{\tau^{2}_{2}+\tau^{2\dagger}_{2}}{\sqrt{3}}~,\qquad\mbox M_{8} = \sqrt{\frac{2}{3}}\tau^{2}_{0}~.
\end{eqnarray}

These operators are explicitly represented in $|1m\rangle$ basis where m = 1, 0, -1  as follows:
\[\fl \eqalign M_{0}= \sqrt{\frac{2}{3}}\left(\begin{array}{cccc}
1 & 0 & 0     \cr
0 & 1 & 0 \cr
0& 0 & 1   \cr
\end{array}\right)\,,\,
M_{1}=\frac{1}{\sqrt{2}}
\left(\begin{array}{cccc}
 0 & -1 & 0   \cr
-1 & 0  & -1  \cr
 0 & -1 &  0  \cr
\end{array}\right)\,,\,
M_{2}=  \frac{i}{\sqrt{2}}
\left(\begin{array}{cccc}
0 & -1 & 0\cr
1 & 0 & -1 \cr
0 & 1 & 0 \cr
\end{array}\right)\,\,,\,\]

\[\fl \eqalign M_{3}= \left(\begin{array}{cccc}
1 & 0 &~~ 0\cr
0 & 0 &~~ 0 \cr
0 & 0 & -1   \cr
\end{array}\right)\,\,,\qquad\mbox
M_{4}= \left(\begin{array}{cccc}
0 &0 & i \cr

0 & 0 & 0 \cr

-i & 0 & 0   \cr
\end{array}\right)\,\,,\qquad\mbox
M_{5}=\frac{i}{\sqrt{2}}
\left(\begin{array}{cccc}
0 & -1 & 0\cr

1 & 0 & 1 \cr

0 & -1 & 0   \cr
\end{array}\right)\,\,,\,\,\]

\[\fl \eqalign M_{6}=\frac{1}{\sqrt{2}}
\left(\begin{array}{cccc}
0 & -1 & 0\cr

-1 & 0 & 1 \cr

0 & 1 & 0   \cr
\end{array}\right)\,\,,\qquad\mbox
M_{7}= \left(\begin{array}{cccc}
0 & 0 & 1\cr

 0 & 0 &0 \cr

 1 & 0 & 0  \cr
\end{array}\right)\,\,,\qquad\mbox
M_{8} = \frac{1}{\sqrt{3}}
\left(\begin{array}{cccc}
1 & 0 & 0\cr

0 & -2 & 0 \cr

0 &0 & 1 \cr
\end{array}\right)\,\,.\,\]

The above matrices are normalized i.e., $Tr(M_{k}M_{k^{'}})= 2\, \delta_{kk^{'}} $ and $M_{1},... , M_{7}$ have eigen values 1, 0, -1.
For a pair of spin 1/2's (two qubits), there are $4\times 4$ linearly independent operators that close under mutual commutator brackets. The 16 
linearly independent operators of a four-state system can be chosen in a variety of matrix representations. One such choice \cite{Van} is 
$\frac {\sigma}{2}$, $\frac {\tau}{2}$, and $\frac {\sigma}{2}\otimes \frac {\tau}{2}$ where $\sigma$ and $\tau$ are the individual pauli matrices 
for two qubits. Together with the $4 \times 4$ unit matrix, these 16 operators can be used to construct any operator describing magnetic couplings 
between the spins as well as the coupling of each spin to an external field. The co-efficient of these operators may in general be functions of 
time and have physical significance. In NMR and quantum computation, the effect of such operator 
combinations on paired spins  and the solutions of the corresponding Hamiltonians is considered.  
Our complete set of basis matrices which are 9 in number are unitarily equivalent to combinations of these operators. i.e.,
\[ \fl  {\sqrt{\frac{2}{3}}}O_{1} \longrightarrow M_{0} \,\,\,, -(O_{5}+O_{9})\longrightarrow M_{1} \,\,\,, (O_{6}+ O_{10})\longrightarrow M_{2} \,\,\,,\]
\[  \fl (O_{2}+ O_{3})\longrightarrow M_{3} \,\,\,, -2 (O_{15}+O_{16})\longrightarrow M_{4} \,\,\,,\,\, 2 (O_{8} + O_{12})\longrightarrow M_{5} \,\,\,,\]
\[ \fl -2 (O_{7}+O_{11})\longrightarrow M_{6} \,\,\,, 2 (O_{13} - O_{14})\longrightarrow M_{7}\,\,\,,\frac {2}{\sqrt {3}}(2O_{4}-O_{13}-O_{14})\longrightarrow M_{8} \,\,.\]

And the unitary transformation which takes an operator from qubit basis to angular momentum basis is 
\[\mathcal{U} =\left(\begin{array}{ccccc}
1 & 0 & 0 & 0 \cr

0 & \frac{1}{\sqrt{2}} & \frac{1}{\sqrt{2}} & 0 \cr

0 & 0 & 0   & 1 \cr

0 & \frac{1}{\sqrt{2}} & -\frac{1}{\sqrt{2}} & 0 \cr
\end{array}\right)\]

As $|1m\rangle $ basis is related to the qubit basis through $|11\rangle$ = $\mid\uparrow\uparrow\rangle$ ,
$|10\rangle$ = $\frac {\mid\uparrow\downarrow\rangle+\mid\downarrow\uparrow\rangle}{\sqrt 2}$, and $|1-1\rangle$ = $\mid\downarrow\downarrow\rangle$ ,
the above 9 matrices in the qubit basis are realized as\\
\begin{eqnarray}
M_0=\sqrt\frac {2}{3}(\mid\uparrow\uparrow\rangle\langle\uparrow\uparrow\mid+\mid\downarrow\downarrow\rangle\langle\downarrow\downarrow\mid) 
 +\frac {1}{6} ((\mid\uparrow\downarrow\rangle+\langle\downarrow\uparrow\mid)+(\langle\uparrow\downarrow\mid+\langle\downarrow\uparrow\mid))\,\,,\nonumber
 \end{eqnarray}
\begin{eqnarray}
M_1  = & -\frac{1}{2}(\mid\uparrow\uparrow\rangle(\langle\uparrow\downarrow\mid+\langle\downarrow\uparrow\mid)+
\mid\uparrow\downarrow\rangle(\langle\uparrow\uparrow\mid+ \langle\downarrow\downarrow\mid)  \nonumber\\
    & + \mid\downarrow\uparrow\rangle(\langle\uparrow\uparrow\mid+\langle\downarrow\downarrow\mid)+
\mid\downarrow\downarrow\rangle(\langle\uparrow\downarrow\mid+\langle\downarrow\uparrow\mid))\,\,,\nonumber
\end{eqnarray}
\begin{eqnarray}
M_2 \, =& \frac{i}{2}(\mid\uparrow\uparrow\rangle(\langle\uparrow\downarrow\mid+\langle\downarrow\uparrow\mid)+
\mid\uparrow\downarrow\rangle(-\langle\uparrow\uparrow\mid+ \langle\downarrow\downarrow\mid)\nonumber\\
      & +\mid\downarrow\uparrow\rangle(-\langle\uparrow\uparrow\mid+\langle\downarrow\downarrow\mid)-
\mid\downarrow\downarrow\rangle(\langle\uparrow\downarrow\mid+\langle\downarrow\uparrow\mid))\,\,,\nonumber
\end{eqnarray}
\begin{eqnarray}
M_3 \,= (\mid\uparrow\uparrow\rangle\langle\uparrow\uparrow\mid)-(\mid\downarrow\downarrow\rangle\langle\downarrow\downarrow\mid)\,\,,\nonumber
\end{eqnarray}
\begin{eqnarray}
M_4\,= i((\mid\downarrow\downarrow\rangle\langle\uparrow\uparrow\mid)-(\mid\uparrow\uparrow\rangle\langle\downarrow\downarrow\mid))\,\,,\nonumber
\end{eqnarray}
\begin{eqnarray}
M_5 \, = & \frac{i}{2}(\mid\uparrow\uparrow\rangle(\langle\uparrow\downarrow\mid+\langle\downarrow\uparrow\mid)-
\mid\uparrow\downarrow\rangle(\langle\uparrow\uparrow\mid+ \langle\downarrow\downarrow\mid)\nonumber\\
      & -\mid\downarrow\uparrow\rangle(\langle\uparrow\uparrow\mid+\langle\downarrow\downarrow\mid)+
\mid\downarrow\downarrow\rangle(\langle\uparrow\downarrow\mid+\langle\downarrow\uparrow\mid))\,\,,\nonumber
\end{eqnarray}
\begin{eqnarray}
M_6 \, = & \frac{1}{2}(-\mid\uparrow\uparrow\rangle(\langle\uparrow\downarrow\mid+\langle\downarrow\uparrow\mid)+
\mid\uparrow\downarrow\rangle(-\langle\uparrow\uparrow\mid+ \langle\downarrow\downarrow\mid)\nonumber\\
      &  +\mid\downarrow\uparrow\rangle(-\langle\uparrow\uparrow\mid+\langle\downarrow\downarrow\mid)+
\mid\downarrow\downarrow\rangle(\langle\uparrow\downarrow\mid+\langle\downarrow\uparrow\mid))\,\,,\nonumber
\end{eqnarray}
\begin{eqnarray}
M_7\,= ((\mid\uparrow\uparrow\rangle\langle\downarrow\downarrow\mid)+(\mid\downarrow\downarrow\rangle\langle\uparrow\uparrow\mid))\,\,,\nonumber
\end{eqnarray}
\begin{eqnarray}
M_8\,= \frac{1}{\sqrt 3}((\mid\uparrow\uparrow\rangle\langle\uparrow\uparrow\mid)-\mid\uparrow\downarrow\rangle(\langle\uparrow\downarrow\mid+\langle\downarrow\uparrow\mid)
-\mid\downarrow\uparrow\rangle(\langle\uparrow\downarrow\mid+\langle\downarrow\uparrow\mid)+\mid\downarrow\downarrow\rangle\langle\downarrow\downarrow\mid)\,\,.\nonumber
\end{eqnarray}

In this representation the most general spin-1 Hamiltonian can be written as 
\begin{equation}
\mathcal H(t) = \frac {1}{2}\sum^{8}_{i=0}\, h_{k}(t)M_{k}\,\,\,.
\end{equation}
Here $M_k$'s in terms of angular momentum operators ${J}_x$,${J}_y$,${J}_z$ are given by 
$M_{1} = -(J_{x})$~,~$M_{2} = (J_{y})$\,,~$M_{3} = (J_{z})$\,,~$M_{4} = -(J_{x}J_{y}+J_{y}J_{x})$~,~$M_{5} = (J_{y}J_{z}+J_{z}J_{y})$~,
$M_{6} = -(J_{x}J_{z}+J_{z}J_{x})$~,~$M_{7} = (J_{x}^{2}-J_{y}^{2})$~,~$M_{8} = (3J_{z}^{2}-2)$\,.
Note that the expansion co-efficients ${h_{k}}= Tr(\mathcal {H}M_{k})$ are real and hence they constitute an experimentally measurable set of parameters.

\subsection {Algebra of Cartesian tensor Operators}
Commutation and anticommutation relations between $M_{k}$'s are given in table I and II respectively.\\
{Table I:} Table of commutators $[M_{k},M_{k{'}}]$, with operators $M_{k}$ in the first column and $M_{k'}$ in the top row, each entry provides the commutator $[M_{k},M_{k{'}}]$.
\begin{equation}
\fl
\begin{array}{cccccccccc}\hline\hline
M_{k} &  M_{1}  &   M_{2} & M_{3} & M_{4} &   M_{5} &  M_{6} &   M_{7} &  M_{8} \cr
\hline
M_{1} &  0      &  -iM_{3} &   iM_{2} & -iM_{6} &  -a & iM_{4}& iM_{5}& \sqrt{3}\,i M_{5} \cr              
M_{2} &  iM_{3} &  0       &   -iM_{1} &  iM_{5} & -iM_{4} & b  &iM_{6} & -\sqrt{3}\,i M_{6}\cr              
M_{3} & -iM_{2} &  iM_{1}  &  0 &  2iM_{7} &  iM_{6}& -iM_{5}& -2iM_{4}& 0  \cr
M_{4} &  iM_{6} &  -iM_{5} &   -2iM_{7}&  0  &  iM_{2}& -iM_{1}& 2iM_{3}& 0   \cr 
M_{5} &   a    &  iM_{4} &   -iM_{6}&  -iM_{2}&  0& iM_{3}& iM_{1}& -\sqrt{3}\,i M_{1}              \cr 
M_{6} & -iM_{4}  & -b  &   iM_{5}&  iM_{1}&  -iM_{3}& 0   & -iM_{2}& \sqrt{3}\,i M_{2}              \cr 
M_{7} & -iM_{5}  & -iM_{6}  &   2iM_{4}&  -2iM_{3} &  -iM_{1}& iM_{2}& 0 & 0              \cr 
M_{8} & -\sqrt{3}\,i M_{5}  &  \sqrt{3}\,i M_{6}  &   0 &  0 &  \sqrt{3}\,i M_{1}& -\sqrt{3}\,i M_{2}  & 0 & 0              \cr 
\hline \hline  \nonumber
\end{array} 
\end{equation}
where $a = i(\sqrt{3}M_{8}+M_{7})$ \,\,, \,\,$b = i(\sqrt{3}M_{8}-M_{7})$.
Here $[M_{3},M_{8}]$=0, $[M_{4},M_{8}]$=0, $[M_{7},M_{8}]$=0. Also there are four vector triplets [$M_{1}, M_{2}, M_{3}$], [$M_{1}, M_{4}, M_{6}$], 
[$M_{4}, M_{2}, M_{5}$], [$M_{5}, M_{3}, M_{6}$],satisfying the relation $[M_{i} , M_{j}] = -i\epsilon_{ijk}M_{k}$
and one triplet [$M_{4}, M_{3}, M_{7}$], satisfying $-2i\epsilon_{ijk}M_{k}$.\\

{Table II:} Table of anticommutators $\{M_{k},M_{k{'}}\}$, with operators $M_{k}$ in the first column and $M_{k'}$ in the top row, each entry provides the anticommutator $\{M_{k},M_{k{'}}\}$.
 \begin{equation}
 \fl
\begin{array}{cccccccccc}
\hline\hline
M_{k}  & M_{1}  &   M_{2} &  M_{3} &  M_{4}  &   M_{5} &  M_{6} &   M_{7}  &  M_{8}    \cr
\hline
M_{1}  &  A  &  M_{4}  &   M_{6}  &  M_{2}  &  0 &  M_{3} & M_{1} & -\frac {1}{\sqrt 3} M_{1}              \cr 
M_{2}  &  M_{4}  &  B   &   M_{5} &  M_{1}   & M_{3}  & 0  & -M_{2} &-\frac{1}{\sqrt3} M_{2}              \cr 
		                                                                 
M_{3}  &  M_{6} &  M_{5}  &  C &  0 &  M_{2} & M_{1}  &  0  &  \frac{2}{\sqrt 3}M_{3}                     \cr 
M_{4}  &  M_{2} &  M_{1} &   0   &  C  & -M_{6}  & -M_{5}  &  0  &    \frac{2}{\sqrt 3}M_{4}                                         \cr 
M_{5}  &  0  &  M_{3}  &   M_{2}   &  -M_{6} &   A   & -M_{4}  & M_{5}    & -\frac{1}{\sqrt 3}M_{5}          \cr 
M_{6}  &  M_{3}  &  0  &   M_{1}  &  -M_{5}  &  -M_{4}  &    B   &  -M_{6} & -\frac{1}{\sqrt 3}M_{6}            \cr 
M_{7}  &  M_{1}  &  -M_{2}  &   0  &  0  &  M_{5}  & -M_{6}  &  C & \frac{2}{\sqrt 3}M_{7}                                             \cr 
M_{8}  &  -\frac {1}{\sqrt 3} M_{1}  &  -\frac {1}{\sqrt 3} M_{2}   &   \frac{2}{\sqrt 3}M_{3}  &  \frac{2}{\sqrt 3}M_{4} &  -\frac {1}{\sqrt 3} M_{5}& -\frac {1}{\sqrt 3} M_{6}  & \frac{2}{\sqrt 3}M_{7} &    D                \cr 
\hline \hline  \nonumber
\end{array} 
\end{equation}
A = $ 2\sqrt{\frac{2}{3}}M_{0}+M_{7}-\frac{1}{\sqrt 3}M_{8}$\,, B = $ 2\sqrt{\frac{2}{3}}M_{0}-M_{7}-\frac{1}{\sqrt 3}M_{8}$\,, C =  $2\sqrt{\frac{2}{3}}M_{0}+\frac{2}{\sqrt 3}M_{8}$\,,
\\ D = $ 2\sqrt{\frac{2}{3}}M_{0}-\frac{2}{\sqrt 3}M_{8}$\,.

The relationship between $M_{k}\,'s$, and the Gell-Mann matrices $\Lambda_{k}'s$ \, k=1...8 is given by

\[\fl M_{1}=-\frac{1}{\sqrt{2}}(\Lambda_{1}+\Lambda_{6}),\, M_{2}=\frac{1}{\sqrt{2}}(\Lambda_{2}+\Lambda_{7}),\,M_{3}=\frac{1}{2}\Lambda_{3}+\frac{\sqrt3}{2}\Lambda_{8}, \,
M_{4}= -\Lambda_{5},\]

\[\fl M_{5}=\frac{1}{\sqrt{2}}(\Lambda_{2}-\Lambda_{7}),\, M_{6}=\frac{1}{\sqrt{2}}(\Lambda_{6}-\Lambda_{1}),\,M_{7}=\Lambda_{4},\,
M_{8}=\frac{\sqrt3}{2}\Lambda_{3}-\frac{1}{2}\Lambda_{8}.\]

\section {Two qubit symmetric gates} 
The algebraic study of $M_{k}'s$ and the corresponding symmetric two qubit gates is important not only for understanding 
fundamental properties of coupled spins and of quantum circuits, but also to study possible experimental implementations in different physical 
systems.  Hamiltonian evolution provides the hardware for quantum gates. i.e., the time evolution of the operators 
$M_{k}$'s provide various symmetric logic gates for quantum computation. The closed form expression for $e^{iM_{k}\theta}$ is given by 
$B_{k}$ = $e^{iM_{k}\theta}$=  $I+(cos{\theta}-1)M_{k}^{2}+isin{\theta}M_{k}$. Here k = 1....7 and I is a $3\times 3$ unit matrix. Following are the 
explicit forms of the gates $B_{k}$'s in the symmetric subspace: 
 
\[\fl B_{1}= \left(\begin{array}{cccc}
cos^{2}\frac {\theta}{2} & \frac {-isin{\theta}}{\sqrt 2} & -sin^{2}\frac{\theta}{2}     \cr

 \frac {-isin{\theta}}{\sqrt 2} & cos\theta & \frac {-isin{\theta}}{\sqrt 2} \cr

 -sin^{2}\frac{\theta}{2}& \frac {-isin{\theta}}{\sqrt 2} & cos^{2}\frac {\theta}{2}   \cr
\end{array}\right),\,\, 
B_{2}= \left(\begin{array}{cccc}
cos^{2}\frac {\theta}{2} & \frac {sin{\theta}}{\sqrt 2} & sin^{2}\frac{\theta}{2}     \cr

 \frac {-sin{\theta}}{\sqrt 2} & cos\theta & \frac {sin{\theta}}{\sqrt 2} \cr

 sin^{2}\frac{\theta}{2}& \frac {-sin{\theta}}{\sqrt 2} & cos^{2}\frac {\theta}{2}   \cr
\end{array}\right),\,\]

\[\fl B_{3}= \left(\begin{array}{cccc}
e^{i\theta} & 0 & 0     \cr

 0 & 1 & 0 \cr

 0& 0 & e^{-i\theta}  \cr
\end{array}\right),
 B_{4}= \left(\begin{array}{cccc}
cos\theta & 0 & -sin\theta    \cr

 0 & 1 & 0 \cr

 sin\theta & 0 & cos \theta   \cr
\end{array}\right),
 B_{5}= \left(\begin{array}{cccc}
cos^{2}\frac {\theta}{2} & \frac {sin{\theta}}{\sqrt 2} & -sin^{2}\frac{\theta}{2}     \cr

 \frac {-sin{\theta}}{\sqrt 2} & cos\theta & \frac {-sin{\theta}}{\sqrt 2} \cr

 -sin^{2}\frac{\theta}{2}& \frac {sin{\theta}}{\sqrt 2} & cos^{2}\frac {\theta}{2}   \cr
\end{array}\right),\,\]

\[\fl B_{6}= \left(\begin{array}{cccc}
cos^{2}\frac {\theta}{2} & \frac {-isin{\theta}}{\sqrt 2} & sin^{2}\frac{\theta}{2}     \cr

 \frac {-isin{\theta}}{\sqrt 2} & cos\theta & \frac {isin{\theta}}{\sqrt 2} \cr

 sin^{2}\frac{\theta}{2}& \frac {isin{\theta}}{\sqrt 2} & cos^{2}\frac {\theta}{2}   \cr
\end{array}\right),
 B_{7}= \left(\begin{array}{cccc}
cos\theta & 0 & isin\theta    \cr

 0 & 1 & 0 \cr

 isin\theta & 0 & cos\theta   \cr
\end{array}\right),
 B_{8}= \left(\begin{array}{cccc}
e^\frac {i\theta}{\sqrt 3} & 0 & 0    \cr

 0 & e^{\frac {-2i\theta}{\sqrt 3}} & 0 \cr

 0& 0 & e^\frac {i\theta}{\sqrt 3}   \cr
\end{array}\right).\]

A useful property of a two qubit symmetric gate is its ability to produce a maximally entangled state from an unentangled one. This property is 
locally invariant.  It is well known that perfect entanglers are those unitary operators that can generate maximally entangled states from some 
suitably chosen separable states. The entangling properties of quantum operators have already been discussed in the literature 
\cite {Zanardi,Balakrishnan,Zhang}. Here we calculate the entangling power of two qubit symmetric gates following the simplified expression given 
by Balakrishnan et al. \cite{Balakrishnan} according to which the gate B is a perfect entangler if its entangling power, $ e_{p}(B)$ = $\frac{2}{9}(1-|G_{1}|)$
has the range $\frac {1}{6}\leq e_{p}\leq \frac{2}{9}$.\\ 
The local invariant $G_{1}$ [Ref. \cite{Makhlin} table II] in terms of symmetric, unitary matrix m is given by $G_{1} = \frac {tr^{2} m}{16 det [B]}$ .
Here m = $B_{B}^{T}B_B$ where the gates in the Bell basis are given by $B_{B} = UBU^{\dagger}$. U is a transformation matrix given by 
\[U =\frac{1}{\sqrt{2}}
\left(\begin{array}{ccccc}
1 & 0 & 1 & 0 \cr

0 & -\sqrt{2}i & 0 & 0 \cr

0 & 0 & 0   & \sqrt{2} \cr

-i & 0 & i & 0 \cr
\end{array}\right)\]
connecting the angular momentum basis $|11\rangle$, $|10\rangle$, $|1-1\rangle$, $|00\rangle$ to the Bell basis 
$\frac {\mid\downarrow\downarrow\rangle+\mid\uparrow\uparrow\rangle }{\sqrt 2}$, $\frac {i(\mid\downarrow\uparrow\rangle+\mid\uparrow\downarrow\rangle)}{\sqrt 2}$,
$\frac {\mid\downarrow\uparrow\rangle-\mid\uparrow\downarrow\rangle}{\sqrt 2}$,$\frac {i(\mid\downarrow\downarrow\rangle-\mid\uparrow\uparrow\rangle)}{\sqrt 2}$.                                          
The relation $ e_{p}(B)$ = $\frac{2}{9}(1-|G_{1}|)$ implies that gates having the same $|G_{1}|$ must necessarily possess the same entangling power 
$e_p$.\\
It is obvious that $B_{1}$, $B_{2}$, $B_{3}$ donot produce entanglement  as they represent rotations which is a local unitary transformation. 
Note that $|G_{1}|$ = 1 and $e_{p}$ = 0 for the above gates. Interestingly, for the gates  $B_{4}$, $B_{5}$, $B_{6}$ and $B_{7}$, $|G_{1}| = Cos^{4}(\theta)$. Observe that since $0\leq G_1\leq 1$ for 
$0\leq\theta\leq\frac{\pi}{2}$, it is clear that $ 0 \leq  e_{p}(B_{B})_{k} \leq \frac {2}{9}$ ( k = 4...7). All these above mentioned gates are 
perfect entanglers for $\frac{\pi}{4}\leq \theta \leq \frac{\pi}{2}$. Similarly the gate $B_{8}$ will have maximum entangling power i.e.,  
$e_p$ = 2/9 when $\theta =  {\sqrt3}\frac{\pi}{2}$. \\  

As an example, consider the  direct product state $|\psi _{12}\rangle$ = $|\psi_{1}\rangle\otimes|\psi_{2}\rangle$, of two spinors in the qubit basis. 
\begin{eqnarray}|\psi _{12}\rangle =  \left(\begin{array}{cc}
 cos\frac {\alpha_{1}}{2} \cr
 sin\frac {\alpha_{1}}{2}e^{i\phi_{1}}\cr
 \end{array}\right)\otimes\left(\begin{array}{cc}
 cos\frac {\alpha_{2}}{2} \cr
 sin\frac {\alpha_{2}}{2}e^{i\phi_{2}}\cr
 \end{array}\right)                  \nonumber \\ 
= \left(\begin{array}{cc}
  cos\frac {\alpha_1}{2}cos\frac {\alpha_2}{2} \cr
 cos\frac {\alpha_1}{2}sin\frac {\alpha_2}{2}e^{i\phi_{2}} \cr
 sin\frac {\alpha_1}{2}cos\frac {\alpha_2}{2}e^{i\phi_{1}}\cr
 sin\frac {\alpha_1}{2}sin\frac {\alpha_2}{2}e^{i(\phi_{1}+\phi_{2})}  \cr\nonumber
 \end{array}\right),\end{eqnarray} 
 $0\leq\alpha_{1,2}\leq\pi$\,, $0\leq\phi_{1,2}\leq2\pi$\,.  Note that a separable state in the symmetric subspace will have the form
 \[|\psi _{12}\rangle_{sym} = \left(\begin{array}{cc}
 cos^{2}\frac {\alpha}{2} \cr
 \sqrt{2} sin\frac {\alpha}{2}cos\frac {\alpha}{2}e^{i\phi} \cr
 sin^{2}\frac {\alpha}{2}e^{2i\phi}\cr
 \end{array}\right),\]\, 
where $\alpha_1$ = $\alpha_2$ = $\alpha$ and $\phi_{1}$ = $\phi_{2}$ = $\phi$. \\
 It is a well known fact that for a pure state of two qubits  $\mid\psi\rangle$ = $ a\mid\uparrow\uparrow\rangle + b\mid\uparrow\downarrow\rangle + 
c\mid\downarrow\uparrow\rangle + d\mid\downarrow\downarrow\rangle$, the 
expression for concurrence is $C(\psi) = 2|ad-bc|$ \cite{Wootters}. For a maximally entangled quantum state concurrence C = 1.
It can be observed that under the action of the  gates $B_{4}$, $B_{7}$ and $B_{8}$ (with $e_{p}$ being maximum i.e., 2/9), 
$|\psi_{12}\rangle_{sym}$ will become maximally entangled state when $\alpha = \frac {\pi}{2}$. i.e.,
\[B_{4}|\psi _{12}\rangle_{sym} \put(-4,8.5){$\alpha=\frac{\pi}{2}$}\longrightarrow \left(\begin{array}{cc}
 -\frac{1}{2}e^{2i\phi}    \cr
 \frac {1}{\sqrt2}e^{i\phi} \cr
 \frac {1}{2}                \cr
 \end{array}\right), 
B_{7}|\psi _{12}\rangle_{sym} \put(-4,8.5){$\alpha=\frac{\pi}{2}$}\longrightarrow \left(\begin{array}{cc}
 \frac{i}{2}e^{2i\phi}    \cr
 \frac {1}{\sqrt2}e^{i\phi} \cr
 \frac {i}{2}                \cr
 \end{array}\right),\] 
 \[B_{8}|\psi _{12}\rangle_{sym} \put(-4,8.5){$\alpha=\frac{\pi}{2}$}\longrightarrow \,\left(\begin{array}{cc}
 \frac{i}{2}    \cr
 -\frac {1}{\sqrt2}e^{i\phi} \cr
 \frac {i}{2}e^{2i\phi}                \cr
 \end{array}\right).\]or in the qubit basis 
\[B_{4}|\psi_{12}\rangle_{sym}\put(-4,8.5){$\alpha=\frac{\pi}{2}$}\longrightarrow -\frac{1}{2}e^{2i\phi}\mid\uparrow\uparrow\rangle +\frac{1}{2}e^{i\phi}\mid\uparrow\downarrow\rangle
+\frac{1}{2}e^{i\phi}\mid\downarrow\uparrow\rangle+\frac{1}{2}\mid\downarrow\downarrow\rangle,\]
\[B_{7}|\psi_{12}\rangle_{sym}\put(-4,8.5){$\alpha=\frac{\pi}{2}$}\longrightarrow\,\, \frac{i}{2}e^{2i\phi}\mid\uparrow\uparrow\rangle +\frac{1}{2}e^{i\phi}\mid\uparrow\downarrow\rangle
+\frac{1}{2}e^{i\phi}\mid\downarrow\uparrow\rangle+\frac{i}{2}\mid\downarrow\downarrow\rangle,\]
\[B_{8}|\psi_{12}\rangle_{sym}\put(-4,8.5){$\alpha=\frac{\pi}{2}$}\longrightarrow -\frac{i}{2}\mid\uparrow\uparrow\rangle +\frac{1}{2}e^{i\phi}\mid\uparrow\downarrow\rangle
+\frac{1}{2}e^{i\phi}\mid\downarrow\uparrow\rangle+\frac{i}{2}e^{2i\phi}\mid\downarrow\downarrow\rangle.\]
Similarly, the gates  $B_{5}$, $B_{6}$ acting on the symmetric separable state transform it into maximally entangled one when
$\alpha = 0, \pi$. For eg:
\[B_{5}|\psi _{12}\rangle_{sym} \put(-4,8.5){$\alpha=0$}\longrightarrow \,\left(\begin{array}{cc}
 \frac{1}{2}       \cr
 -\frac {1}{\sqrt2} \cr
 -\frac {1}{2}       \cr
 \end{array}\right), 
 B_{6}|\psi _{12}\rangle_{sym} \put(-4,8.5){$\alpha=0$}\longrightarrow\,\left(\begin{array}{cc}
 \frac{1}{2}           \cr
 -\frac {i}{\sqrt2} \cr
 \frac {1}{2}                \cr
 \end{array}\right).\] 
\[B_{5}|\psi_{12}\rangle_{sym}\put(-4,8.5){$\alpha=0$}\longrightarrow \frac{1}{2}\mid\uparrow\uparrow\rangle -\frac{1}{2}\mid\uparrow\downarrow\rangle
-\frac{1}{2}\mid\downarrow\uparrow\rangle-\frac{1}{2}\mid\downarrow\downarrow\rangle.\]
\[B_{6}|\psi_{12}\rangle_{sym}\put(-4,8.5){$\alpha=0$}\longrightarrow \frac{1}{2}\mid\uparrow\uparrow\rangle -\frac{i}{2}\mid\uparrow\downarrow\rangle
-\frac{i}{2}\mid\downarrow\uparrow\rangle+\frac{1}{2}\mid\downarrow\downarrow\rangle.\]
It can be noted that the operators $B_{8}$ and $B_{4}$  produce spin squeezing resulting from a single axis twisting and two axis counter twisting 
respectively \cite{Kitagawa}. Also possibility of physical realization of these spin squeezing operators are given in Ref.\cite{Pathak}.

\subsection{Special perfect entanglers} Rezakhani \cite{Rezakhani} has analyzed the perfect entanglers and found that some of them have the unique 
property of maximally entangling a complete set of orthonormal product vectors. Such operators for which $e_{p} = \frac{2}{9}$ belong to a well 
known family of special perfect entanglers. A study of using such special perfect entanglers as the building blocks of the most efficient universal gate simulation is also given 
in ref.\cite{Rezakhani}. Let us now study the conditions under which the perfect entanglers can be classified as special perfect entanglers. 
When $e_{p} = \frac{2}{9}$, $B_{4},.......B_{8}$ in the qubit basis are given by  
\[ B_{4}= \left(\begin{array}{ccccc}
0 & 0 & 0 & -1    \cr

 0 & 1 & 0  & 0   \cr

 0 & 0 & 1 & 0    \cr
 
 1 & 0 & 0 & 0   \cr
\end{array}\right)\,,\,\,
B_{5}= \frac{1}{2}\left(\begin{array}{ccccc}
1 & 1 & 1 & -1    \cr

 -1 & 1 & -1  & -1  \cr

 -1 & -1 & 1 & -1    \cr
 
 -1 & 1 & 1 & 1   \cr
\end{array}\right)\,,\,\,\]
\[\fl B_{6}= \frac{1}{2}\left(\begin{array}{ccccc}
1 & -i & -i & 1    \cr

 -i & 1 & -1  & i  \cr

 -i & -1 & 1 & i    \cr
 
 1 & i & i & 1   \cr
\end{array}\right)\,,\,\,
B_{7}= \left(\begin{array}{ccccc}
0 & 0 & 0 & i    \cr

 0 & 1 & 0  & 0   \cr

 0 & 0 & 1 & 0    \cr
 
 i  & 0 & 0 & 0   \cr
\end{array}\right)\,,\,\,
B_{8}= \left(\begin{array}{cccc}
i & 0 & 0 & 0    \cr

 0 & 0 & -1  & 0   \cr

 0 & -1 & 0 & 0    \cr
 
 0  & 0 & 0 & i   \cr
\end{array}\right)\,\,.\,\,\]
Following Rezakhani \cite{Rezakhani}, the most general separable basis (upto general phase factors for each vector) is 

\[|\psi_{1}\rangle = (a|\uparrow\rangle+b|\downarrow\rangle)\otimes(c|\uparrow\rangle+d|\downarrow\rangle)\,,\]        
\[|\psi_{2}\rangle = (-b^{*}|\uparrow\rangle+a^{*}|\downarrow\rangle)\otimes(c|\uparrow\rangle+d|\downarrow\rangle)\,,\]  
\[|\psi_{3}\rangle = (e|\uparrow\rangle+f|\downarrow\rangle)\otimes(-d^{*}|\uparrow\rangle+c^{*}|\downarrow\rangle)\,,\]   
\[|\psi_{4}\rangle = (-f^{*}|\uparrow\rangle+e^{*}|\downarrow\rangle)\otimes(-d^{*}|\uparrow\rangle+c^{*}|\downarrow\rangle)\,,\]
where $|a|^{2}+|b|^{2}= |c|^{2}+|d|^{2}= |e|^{2}+|f|^{2}$ = 1. 

When the gates $B_{4}$, $B_{7}$ and $B_{8}$ as perfect entanglers act on the state - say $|\psi_{1}\rangle$, we obtain 
\[[B_{4,7,8}]|\psi_{1}\rangle = -bd\,|\uparrow\uparrow\rangle+ad\, |\uparrow\downarrow\rangle+bc\, |\downarrow\uparrow\rangle+ac\, |\downarrow\downarrow\rangle. \]
This state is maximally entangled if its concurrence, C = $4|abcd|$ = 1. Thus these two qubit symmetric gates transform the orthonormal states 
$|\psi_{1}\rangle$, $|\psi_{2}\rangle$, $|\psi_{3}\rangle$ and $|\psi_{4}\rangle$ into maximally entangled ones if $|abcd| = |cdef| = \frac {1}{4}$. 
Similarly, for the gates $B_{5}$ and $B_{6}$, condition for finding a full set of  orthonormal product states is 
$|(a^{2}+b^{2})(c^{2}+d^{2})| = |(e^{2}+f^{2})(c^{2}+d^{2})|=1 $.    

Let us consider the example of Lipkin-Meshkov-Glick interaction Hamiltonian \cite{Lipkin, Pathak} which is widely used in nuclear physics. 
\begin{equation}
\mathcal {H}_{L} = \mathcal {G}_{1}(J^{2}_{+}+J^{2}_{-})+\mathcal {G}_{2}(J_{+}J_{-}+J_{-}J_{+})\,\,.
\end{equation}
Here $\mathcal {G}_{1}$ and $\mathcal {G}_{2}$ are the coupling constants. In terms of our operators $M_{k}\,'s$, 
\begin{equation}
 \mathcal {H}_{L} = \mathcal {G}_{1}^{\prime} M_{7} + \mathcal {G}_{2}^{\prime}(\sqrt {8}M_{0} -M_{8})\,\,,
\end{equation}
where $\mathcal {G}_{1}^{\prime} = 2\mathcal {G}_{1}$ and $\mathcal {G}_{2}^{\prime} = \frac{2}{\sqrt 3}\mathcal {G}_{2}$. Since $[M_{7},M_{8}]$ = 0, we have
\[ e^{iH_{L}t} = B_{L}= \left(\begin{array}{cccc}
e^{\sqrt {3}\,i\beta} cos\xi & 0 & ie^{\sqrt {3}\,i\beta}sin\xi     \cr

 0 & e^{2\sqrt {3}\,i\beta}   & 0 \cr

 ie^{\sqrt {3}\,i\beta} cos\xi & 0 & e^{\sqrt {3}\,i\beta} cos\xi    \cr
\end{array}\right),\,\,\]
in spin-1 subspace. Here $\xi = \mathcal {G}_{1}^{\prime}t$ and  $\beta = \mathcal {G}_{2}^{\prime}t$ and $e_{p}= \frac{2}{9}$ for
$2\mathcal {G}_{2} t = \frac {\pi}{2}+ 2 \mathcal {G}_{1} t$.  Under the action of this gate (with $e_{p} = \frac {2}{9}$), the separable state 
$|\uparrow\uparrow\rangle (|\downarrow\downarrow\rangle)$ becomes entangled for all values of t except when t = $\frac{n\pi}{4\mathcal {G}_{1}}$; 
n=0,1,2...... and maximally entangled when $4\mathcal {G}_{1}t$ = $(2n+1)\frac{\pi}{2}$.   
For eg., \[B_{L}|\psi_{12}\rangle_{sym}\put(-4,8.5){$\alpha=0$}\longrightarrow cos(2\mathcal{G}_{1}t)\mid\uparrow\uparrow\rangle + isin(2\mathcal{G}_{1}t)\mid\downarrow\downarrow\rangle\,\,.\]
\section{Conclusion}
In conclusion, we have constructed traceless,Hermitian and linearly independent set of basis matrices which provides an alternative representation
of SU(n) generators. We have considered unitary evolutions of two spin-1/2 states in angular momentum subspace (j=1) and constructed physically 
realizable logic gates using (2j+1) dimensional  representation of the above set of basis matrices. Entangling properties of these gates have been 
studied in terms of their entangling power $e_{p}$. $e_{p}$ is found to be maximum (2/9) for $B_{4}, ..., B_{8}$ under certain conditions which is the signature for 
special perfect entanglers. These logic gates are obtained by the exponentiation of the quadratic form of angular momentum operators $\bf{J}_{x},
\bf{J}_{y},\bf{J}_{z}$.
As an example we have taken the well known Lipkin-Meshkov-Glick Hamiltonian and studied its entangling properties in spin-1 subspace. Further, we have shown that precisely 
at what time the initial separable state becomes maximally entangled under the action of  perfect entanglers 
which consists of one-axis twisting and two axis twisting Hamiltonians that produce spin squeezing.

\ack     
One of us (V.A.) acknowledges with thanks the support provided by the University Grants Commission (UGC), India for
the award of teacher fellowship through Faculty Development Programme (FDP).

\end{document}